\documentstyle[12pt]{article}
\begin{document}
\title{A very brief review of Bose-Einstein correlations}
\author{Kacper Zalewski\thanks{Supported in part by the KBN grant
2P03B 086 14}\\
M. Smoluchowski Institute of Physics, Jagellonian University\\ and\\ Institute
of Nuclear Physics, Krak\'ow, Poland}
\maketitle

\begin{abstract}
The GGLP approach to Bose-Einstein correlations, a hydrodynamic model and a
string model are briefly reviewed. The implications of the two models for the
Bose-Einstein correlations among the decay products of a pair of $W$ bosons are
presented.
\end{abstract}
\maketitle

\section{INTRODUCTION}

Several reviews of Bose-Einstein correlations have been recently published
\cite{WIH,WEI,HEJ}. The two reviews published in Physics Reports are each about
hundred pages long and are selective. The overlap between them is not large. In
order to give a review in the 20 minutes allotted to me, I will have to be very
selective. Let me first give three reasons, why the Bose-Einstein correlations
are considered to be a good research problem.

Experimentally the effect is spectacular. The experimentalists, coauthors of
the seminal paper \cite{GGL} (further quoted as GGLP) failed to discover the
$\rho^0$ meson. This may seem strange to people used to present day
experiments, because the $\rho^0$ is the most conspicuous resonance in the
$\pi^+\pi^-$ system, but at that time statistics was not good enough.
Nevertheless, the Bose-Einstein correlations were clearly seen. The
distributions of the opening angles between the momenta for pairs of like-sign
pions ($\pi^+\pi^+$ or $\pi^-\pi^-$) were significantly shifted towards smaller
angles as compared to the corresponding distribution for the unlike-sign pions
($\pi^+\pi^-$).

Bose-Einstein correlations in multiple particle production processes give, in
principle, access to quantitative information about the structure of the source
of hadrons, such as the geometrical size of the source and its shape, the life
time of the source, the fraction of this lifetime, when the hadrons are
actually being produced etc. There is no other quantitative method of obtaining
this information \cite{WIH}.

The problem is hard. At the Marburg conference in 1990, G. Goldhaber, one of
the creators of this field of research, said: "What is clear is that we have
been working on this effect for thirty years. What is not as clear is that we
have come much closer to a precise understanding of the effect". That was
Goldhaber's opinion ten years ago. What would be his statement now? What is
clear is that he would have changed thirty years to forty years. What is not
as clear is that he would have considered any other changes necessary. Whether
or not one agrees with Goldhaber that the GGLP paper contains in the nut shell
all we understand about the Bose-Einstein correlations, there is no doubt that
this is a very important paper and that without knowing the main results
contained there it is not possible to discuss Bose-Einstein correlations in
multiple particle production processes. Therefore, we will briefly review these
results in the following section.

Let us conclude this introduction by a remark on terminology. For some years
after the discovery, the Bose-Einstein correlations in multiple particle
production processes were known as the GGLP effect. In the seventies it was
pointed out by a number of people that the interference of intensities, basic
for the GGLP effect, had been used earlier by astronomers to determine the
radii of stars. From the names of Hanbury Brown and Twiss, who introduced this
method in astronomy, the effect was renamed the HBT effect. This name is still
popular, but since the effect results from standard Bose-Einstein statistics,
the name BEC, for Bose-Einstein correlations, is now gaining ground.

\section{THE GGLP RESULTS}

Consider two $\pi^+$-mesons produced one at point $\vec{r}_1$, the other at
point $\vec{r}_2$. Let the momenta of the two pions be $\vec{p}_1$ and
$\vec{p}_2$ respectively. Suppose first that the two positive pions are
distinguishable. Then it makes sense to assume that the pion with momentum
$\vec{p}_1$ originated at $\vec{r}_1$ and that with momentum $\vec{p}_2$ at
$\vec{r}_2$. A reasonable expression for the probability amplitude to observe
the two pions at $\vec{r}$ would be

\begin{equation}
A_D = e^{i\phi_1 + i\vec{p}_1\cdot(\vec{r} - \vec{r}_1)}e^{i\phi_2 +
i\vec{p}_2\cdot(\vec{r} - \vec{r}_2)}.
\end{equation}

This is the product of two single-particle amplitudes for the two independently
produced pions. Each amplitude has a phase, which is the sum of the phase
obtained by the pion at birth and of the phase accumulated by the pion in the
process of propagation with its momentum from its birth point to point
$\vec{r}$. Since the two pions are identical particles, this amplitude is
unacceptable. A reasonable amplitude must be symmetric with respect to
exchanges of the two pions. Thus, for two identical pions the amplitude
corresponding to amplitude (1) is

\begin{eqnarray}
A = & \frac{1}{ \sqrt 2}
 e^{i( \phi_ + \phi_2) + i( \vec{p}_1+\vec{p}_2) \cdot\vec{r}}*\nonumber \\
 &
\!\!\!\!\!\!\!\!\!\!\!\!\!\!\!\!\left(e^{-i(\vec{p}_1\cdot\vec{r}_1+\vec{p}_2\cdot\vec{r}_2)}
+
 e^{-i( \vec{p}_2\cdot\vec{r}_1+\vec{p}_1\cdot\vec{r}_2)}\right).
\end{eqnarray}
The points $\vec{r}_1$ and $\vec{r}_2$ are not known, therefore, for comparison
with experiment this amplitude should be averaged over some assumed
distribution of these points in space. If this is done, nothing interesting
is obtained, and here comes the first brilliant idea of GGLP: they assumed that
what should be averaged is the squared modulus of this amplitude:

\begin{equation}
\mid A\mid^2 = 1 + \cos\left[
(\vec{p}_1-\vec{p}_2)\cdot(\vec{r}_1-\vec{r}_2)\right].
\end{equation}
Physically this assumption means that the contributions from various pairs of
points $(\vec{r}_1,\vec{r}_2)$ add incoherently. The average

\begin{eqnarray}
\langle|A|^2\rangle = 1 +& \nonumber \\ \int d^3r_1 d^3r_2&\!\!\!\!\!
\rho(\vec{r}_1,\vec{r}_2)\cos\left[
(\vec{p}_1-\vec{p}_2)\cdot(\vec{r}_1-\vec{r}_2)\right]
\end{eqnarray}
depends, of course, on the assumed distribution of the points of origin
$\rho(\vec{r}_1,\vec{r}_2)$. GGLP chose

\begin{equation}
\rho( \vec{r}_1,\vec{r}_2) = (2\pi R^2)^{-3}\exp\left[-\frac{r_1^2 +
r_2^2}{2R^2}\right],
\end{equation}
where $R$ is a parameter with the dimension of length, which may be
interpreted as the radius of the source of hadrons, and obtained

\begin{eqnarray}
\langle\mid A\mid^2\rangle = 1 + e^{-\vec{q}^2R^2},
\end{eqnarray}
where

\begin{equation}
\vec{q} = \vec{p}_1 - \vec{p}_2.
\end{equation}
This function of $\vec{q}^2$ drops from the value two at $\vec{q}^2 = 0$ to
values close to one for $\vec{q}^2$ large. The width of the forward
peak is of the order of $R^{-2}$. Note that these results are qualitatively
much more general than the Gaussian Ansatz used by GGLP. For $\vec{q}^2 = 0$
the cosine is equal one and so is its average value. For large values of
$\vec{q}^2$ the cosine is a rapidly oscillating function of $(\vec{r}_1 -
\vec{r}_2)$ and its average with any smooth distribution is very small. If the
distribution contains only one parameter with dimension of length $R$, the
width of the forward peak must be of order $R^{-2}$ for dimensional reasons.
In spite of this generality the GGLP formula for the two-particle distribution
is unrealistic. It is certainly not true that the probability of producing
a pair of pions depends only on their momentum difference. Here comes the
second brilliant idea of GGLP. They noticed that in their model the two
particle distribution for distinguishable pions is a constant equal one.
Consequently, their result can be reinterpreted as

\begin{equation}
\label{ratio}
\frac{\rho_2(\vec{p}_1,\vec{p}_2)}{\rho_{2D}(\vec{p}_1,\vec{p}_2)} =
 1 + e^{-\vec{q}^2R^2},
\end{equation}
where $\rho_2$ is the actual distribution of momenta for pairs of identical
pions and $\rho_{2D}$ is the corresponding distribution, if there were no
Bose-Einstein correlations. The problem is that there is no experiment, where
the production of distinguishable $\pi^+$-mesons could be studied and the
distribution $\rho_{2D}$ measured.  The question how to cope with this
difficulty has been extensively discussed without a clear conclusion. For a
(pessimistic) review cf. \cite{HAY}. GGLP chose to put for $\rho_{2D}$ the
distribution for $\pi^+\pi^-$ pairs, where there are (almost) no Bose-Einstein
correlations. They found agreement with experiment and derived a plausible
value for the radius $R$.

Since in the GGLP analysis one talks about pions produced at given points with
given momenta, it may seem inconsistent with quantum mechanics. This is,
however, not the case, because one may obtain the GGLP results starting with
the single particle density matrix

\begin{eqnarray}
\rho_1(\vec{p},\vec{p'}) =
\langle\vec{p}\mid\vec{r}\rangle\rho(\vec{r})\langle\vec{r}\mid\vec{p'}\rangle,
\end{eqnarray}
where

\begin{eqnarray}
 \rho( \vec{r})=(2\pi R^2)^{-3/2}\exp\left[-\frac{r^2}{2 R^2}\right],
\end{eqnarray}
which is legal in quantum mechanics, and obtain the two-particle distribution
by symmetrizing the product of two such density matrices and taking the
diagonal elements

\begin{eqnarray}
\rho_2(\vec{p}_1,\vec{p}_2) =
\rho_1(\vec{p}_1,\vec{p}_1)\rho_1(\vec{p}_2,\vec{p}_2) + \nonumber \\
 \rho_1(\vec{p}_1,\vec{p}_2)\rho_1(\vec{p}_2,\vec{p}_1).
\end{eqnarray}
Note that in these formulae $\rho_1$ is a single particle density matrix, while
$\rho_2$ is a two-particle momentum distribution. The GGLP analysis,
nevertheless, has the defect that it leads to ratios (\ref{ratio}), which
depend only on the difference of momenta, which experimentally is not quite
true.

In the following we discuss two more recent models of Bose-Einstein
correlations in multiple particle production processes. Both are good and
respectable models, but they are based on completely different physical
assumptions, which nicely illustrates, how basic are the problems still
controversial in this branch of physics.

\section{A HYDRODYNAMIC MODEL}

A convenient starting point for models of the kind discussed in this section is
the source function $S(X,K)$ \cite{GKF,PRA} related to the single particle
density matrix by the formula

\begin{eqnarray}
\label{sourcf}
\rho(\vec{p},\vec{p'}) = \int d^4X e^{iqX}S(X,K).
\end{eqnarray}
In this formula

\begin{equation}
q = p - p'
\end{equation}
and

\begin{eqnarray}
K = (p + p')/2
\end{eqnarray}
are fourvectors. Formula (\ref{sourcf}) is similar to the well-known formula
relating the Wigner function $W(\vec{X},\vec{K})$ to the density matrix

\begin{equation}
\rho(\vec{p},\vec{p'}) = \int d^3X e^{-i\vec{q}\cdot\vec{X}}W(\vec{X},\vec{K}).
\end{equation}
Therefore, the source function is often called a kind of Wigner function, a
generalized Wigner function, a pseudo-Wigner function etc.  In fact the
relation of this function to the Wigner function is rather complicated. One
could, of course put

\begin{eqnarray}
S(X,K) = W(\vec{X},\vec{K})\delta(X_0),
\end{eqnarray}

\noindent but since there is an infinity of different source functions giving
the same density matrix, this is not the only choice. Model builders try to
guess a good source function interpreting it as a space time ($X$), momentum
($K$) distribution of sources of hadrons. For instance, if the pions are
produced by classical sources, one can show \cite{SHU,CHH} that

\begin{eqnarray}
S(X,K) = & \nonumber \\
\int \frac{d^4y}{2(2\pi)^3}e^{-iKy}&\!\!\!\!\!\langle
J^*(X+y/2)J(X-y/2)\rangle,
\end{eqnarray}

\noindent where $\langle ... \rangle$ denotes averaging over the distributions
of sources in various events. Whether the source function is guessed, or
derived from some model, it can be used to find the density matrix in the
momentum representation according to formula (\ref{sourcf}). One should be
careful to choose only such source functions, which give distributions of
momenta and positions consistent with Heisenberg's uncertainty principle. In
practice this usually imposes mild limitations on the parameters of the model.

As an illustrative example let us consider the model described in ref.
\cite{WIH} and in more detail in the references quoted there. The source
function is assumed in the form

\begin{eqnarray}
\label{hydsou}
S(X,K) = C m_T \cosh(y - \eta)* \nonumber \\
\exp\left[-(m_T\cosh y \cosh \eta_t - \frac{xK_T}{r_T}\sinh
\eta_t)/T\right]*\nonumber\\
\exp\left[-\frac{r_T^2}{2 R^2} - \frac{\eta^2}{2(\Delta\eta)^2} - \frac{(\tau -
\tau_0)^2}{2 (\Delta\tau)^2}\right].
\end{eqnarray}
This formula is tailored for central collisions of heavy ions. The $Z$-axis is
along the beam direction and includes the centres of the two colliding nuclei.
The $X$ axis is chosen so that the momentum $\vec{K}$ is in the $Z,X$ plane.
The variables connected to time $t$ and to the longitudinal motion are: the
centre of mass rapidity $y$, the pseudorapidity

\begin{eqnarray}
\eta = \frac{1}{2}\ln \frac{t+z}{t-z}
\end{eqnarray}

\noindent and the longitudinal proper time

\begin{eqnarray}
\tau = \sqrt{t^2 - z^2}.
\end{eqnarray}

\noindent The variables related with the transverse motion are: the transverse
component of the momentum vector $\vec{K}$, $K_T = |K_X|$, the
transverse mass

\begin{equation}
m_T = \sqrt{m^2 + K_T^2},
\end{equation}
the distance from the $z$-axis $r_T = \sqrt{X^2 + Y^2}$ and the transverse
flow rapidity

\begin{eqnarray}
\eta_t(r_T) = \eta_f \frac{r_T}{R}.
\end{eqnarray}

\noindent The free parameters of the model are $C,T,R$, $\Delta\eta$,
$\Delta\tau$, $\tau_0$ and $\eta_f$. These parameters have simple physical
interpretations related to the physical interpretation of the Ansatz
(\ref{hydsou}). The exponent in the third line of formula (\ref{hydsou})
implies that the particles are produced not too far from the $z$ axis, a
typical distance being $R$, with rapidities not too far from the centre of
mass rapidity (chosen as equal zero), a typical rapidity being $\Delta\eta$,
and at longitudinal proper time not too far from $\tau_0$, a typical proper
time shift with respect to $\tau_0$ being $\Delta\tau$. The exponential in the
second line is a Boltzmann factor, evaluated in the rest frame of a fluid
element flowing with longitudinal rapidity $y$ and radially with a transverse
rapidity $\eta_t(r)$. T is the temperature and the transverse rapidity is
parametrized in terms of the parameter $\eta_f$.

This model has been fitted to the data of the NA49 experiment for Pb-Pb central
collision at laboratory energy of 158GeV per nucleon \cite{WIH}. The results
are

\begin{eqnarray}
R & \approx & 7 \mbox{ fm},\\
T & \approx & 130 \mbox{ MeV}, \\
\eta_f & \approx & 0.35, \\
\Delta\eta & \approx & 1.3,\\
\tau_0 & \approx & 9 \mbox{ fm},\\
\Delta\tau & \approx & 1.5 \mbox{ fm}.
\end{eqnarray}
Some of these numbers have interesting physical implications. Thus, the radius
$R$ is about twice larger than the radius expected from the known size of the
lead nucleus. This indicates a significant radial expansion before the
hadronization process. The temperature $T$ is lower than temperatures obtained
in models used to calculate the chemical composition of the produced hadrons.
This may be interpreted as evidence, that the transverse expansion is
accompanied by cooling. The value of $\eta_f$ corresponds to velocities of the
order of the sound velocity in a plasma, which confirms that the model is
reasonable. The small ratio $\Delta\tau/\tau_0$ indicates that the
hadronisation process is short compared to the proper time between the
collision and the beginning of hadronization, or freeze out as it is called.
This last conclusion should be taken with care, because the authors stress that
the value of $\Delta\tau$ is poorly constrained by the data. Thus within a
given model it is possible to obtain much valuable, quantitative information.
Much less is known on how model dependent such results are.

A much discussed problem is, whether there should be much difference between
the Bose-Einstein correlations for pairs of identical pions produced in the
decay of a $W$ boson and for the pairs of identical pions, where each of the
pions originates from the decay of a different $W$ boson. Experimentally this
is a problem for $e^+e^-$ annihilations in LEP200. Since the experimental
situation is not clear it is interesting to know, what are the theoretical
predictions of various models. The model discussed in the present section has
not been adapted to $e^+e^-$ annihilations, but for any model with a similar
philosophy there should be little or no difference between the two kinds of
pairs, provided the two $W$ bosons are sufficiently close to each other in
ordinary space and in momentum space, which seems to be the case in LEP200.

\section{A STRING MODEL}

There is a variety of string models. The model discussed here
\cite{ANH,ANR,AND} is an extension of the LUND model. It is tailored for
$e^+e^-$ annihilations with hadron production. When an electron and a positron
collide, usually a quark antiquark pair is first formed. The two partons fly
away from each other with the speed of light (we neglect quark masses at this
stage). The colour field is confined to a string with the quark at one end and
the antiquark at the other. Forming a colour string, however, costs energy. The
energy in the string is

\begin{equation}
E=\kappa L,
\end{equation}

\noindent where L is the length of the string and $\kappa \approx 1$GeV/fm is a
constant known as the string tension. Let us choose the $z$-axis in the
direction of the centre of mass momentum of the quark. Then the antiquark flies
in the $-z$ direction and the string stretches (approximately) along the
$z$-axis. At some moment there is no more energy to extend further the string
and the directions of flight of the two partons get reversed. There is another
interesting possibility. A piece of the string of length $\delta x$ (away from
the ends of the string) can disappear and the energy $\kappa \delta x$ released
in the process can be converted into the transverse masses of the quark and
antiquark formed at the new ends of the strings. From energy and momentum
conservation the two new partons have equal and opposite transverse momenta
related to $\delta x$ by the relation $\kappa \delta x = 2 m_T$. In the
language of quantum theory producing such a break means producing two opposite
(colour) charges at a distance $\delta x$ along the $z$-axis in a uniform force
field also along the $z$ axis. The probability decreases rapidly with increasing
$\delta x$. Therefore, the probability of producing a parton with a transverse
momentum $k_T$ also decreases with increasing $k_T$. One finds

\begin{equation}
\label{ptrans}
P(k_T) \sim \exp\left(-\frac{k_T^2}{2 \sigma^2}\right),
\end{equation}

\noindent where $\sigma^2$ is a constant of order $\kappa/(2\pi)$. A meson is
produced, when a quark and an antiquark forming the ends of a bit of string
meet and inherits their transverse momenta. Thus the model explains, why the
transverse momenta of hadrons are small.

The description of the longitudinal motion is more complicated. Of great
importance is a contour in the $z,t$ plane constructed as follows. One can
assume that all the quarks and antiquarks move with the velocity of light $\pm
c$ $(c=1)$ and consequently in a $(z,t)$ plot travel along lines making angles
of $\pm\pi/4$ with the $z$-axis. Some move to the right (increasing $z$),
others to the left (decreasing $z$). Suppose that the original quark moves to
the right and up, (increasing time). At some point it turns and begins moving
to the left and up. At some other point it meets an antiquark and forms a
hadron (this is an oversimplification, but harmless in the present context).
The antiquark, however, must have been moving to the right and up, in order to
have met the quark moving to the left and up. Following the antiquark line of
motion in the opposite direction (or assuming after Feynman that the antiquark
moves backward in time) we continue the line from the point, where the hadron
was formed, to the point, where the antiquark was formed, i.e. to the point,
where the string broke. From this point, we can continue along the line going
to the left and up following the quark produced together with our antiquark and
reach in this way another point, where a hadron was formed. Continuing this
procedure, we can ascribe to each event a closed contour formed of straight
sections connecting first the point, where the original quark-antiquark pair
was created in the $e^+e^-$ annihilation to the point, where the quark reversed
its direction of motion, then from this point to the point, where it
hadronized, from this point to the point, where the antiquark involved in the
hadronization process was created in a process of string breaking, from there
to the point, where the quark produced in this string breaking process
hadronized and so on until we reach the point, where the original antiquark
changed its direction and finally again the point, where it had been created in
the $e^+e^-$ annihilation. On the $(z,t)$ plane this contour encloses a surface
of area $A$. The crucial assumption is that the probability amplitude for the
event is proportional to

\begin{equation}
\label{stramp}
Ampl(A) = \exp{i\xi A},
\end{equation}
where

\begin{equation}
\xi = \kappa + \frac{ib}{2}
\end{equation}
is a complex constant. The imaginary part of $\xi$ introduces into the
probability of the process the factor

\begin{equation}
|Ampl(A)|^2 = \exp(-bA),
\end{equation}

\noindent which is well known from the standard Lund model. For the description
of the Bose-Einstein correlations, however, it is the real part of $\xi$, which
is important. Note that this model, as opposed to the one described in the
previous section and in the GGLP approach, does not introduce random phases or
explicit assumptions about incoherence. There is no rigorous proof of the basic
relation (\ref{stramp}), but there are strong plausibility arguments in favour
of it \cite{ANH,ANR,AND} --- one based on the properties of Wilson loops, and
another on the WKB approximation.

In order to evaluate the effects of Bose-Einstein correlations, one proceeds in
the standard way. The amplitude for producing $n$ particles with given momenta
is replaced by the sum of all the amplitudes differing only by exchanges of
identical particles. When two identical particles are exchanged, the contour
described above changes. Let us denote the area enclosed by the contour
corresponding to the permutation $P$ of the identical particles by $A_P$. We
will further use the notation

\begin{equation}
\Delta A_{PP'} = A_P - A_{P'}.
\end{equation}

Another effect is that in order to recover the sum zero for the transverse
momenta of the two partons produced at each point, where the string breaks,
some changes in the transverse momenta must accompany an exchange of identical
particles. We will need the change in the sum of squares of the transverse
momenta $(\Delta\sum k_T^2)_{PP'}$. Because of the changes in the area $A$ and
in the sum of the squares of transverse momenta, the probability amplitude for
the production process changes, when identical particles are permuted. Let us
denote the amplitude corresponding to permutation $P$ by $M_P$, and the total
amplitude by $M$. Then

\begin{equation}
M = \sum_P M_P
\end{equation}
and the probability of a given final state is proportional to the squared
modulus of this amplitude, which can be written in the form

\begin{equation}
|M|^2 = \sum_P |M_P|^2w_P,
\end{equation}
where the weight factors

\begin{eqnarray}
w_P = 1 + \sum_{P'\neq P} \frac{2 Re(M_P M_{P'}^*)}{|M_P|^2 + |M_{P'}|^2}.
\end{eqnarray}

\noindent The approximation $w_P = 1$ for all permutations $P$, corresponds to
the case of no Bose-Einstein correlations. In general, using formulae
(\ref{ptrans}) and (\ref{stramp}) one finds

\begin{eqnarray}
w_p = 1 + \sum_{P' \neq P} \frac{\cos\frac{\Delta
A_{PP'}}{2\kappa}}{\cosh\left(
\frac{b\Delta A_{PP'}}{2} + \frac{(\Delta \sum k_T^2)_{PP'}}{4\sigma^2}
\right)}.
\end{eqnarray}

A contribution $PP'$ to the sum on the left is important only, if the change of
the area $A$, when going from permutation $P$ to permutation $P'$, is not big
as compared to $b^{-1}$. An inspection of the contour shows, that such pairs of
permutations differ by exchanges of pairs of particles close to each other
along the contour. This in turn are exchanges of particles with production
points close to each other and, due to the strong correlation of the production
point and momentum in the string model, exchanges of particles with similar
momenta. This agrees with GGLP and with experiment: for pairs of identical
particles the strongest effect is observed, when the momenta of the two
particles are similar. In general the string picture gives very reasonable
results. What is interesting, however, is that it also predicts some new
effects, which do not follow from GGLP or from the hydrodynamic picture.

Compare a pair of $\pi^0$ mesons and a pair of $\pi^+$ meson. In the other
models the effect of Bose-Einstein symmetry in the two cases should be similar.
The analysis requires some care, as the corrections for Coulomb and strong
interactions in the final state are different, but the Bose-Einstein
correlations are handled in the same way. In the string model, however, one
easily notices, that it is not possible to produce two $\pi^+$ mesons next to
each other along the contour. More generally, no segment of the contour can
contain an exotic combination of hadrons. On the other hand, two $\pi^0$ mesons
can be produced next to each other. Consequently \cite{ANR}, the effect of
Bose-Einstein correlations for $\pi^0\pi^0$ pairs is expected to be stronger
than that for $\pi^+\pi^+$ pairs and the effective radius of the hadronization
region $R$ calculated from $\pi^0\pi^0$ pairs should be smaller than that for
$\pi^+\pi^+$ pairs. These effects are not very strong and the measurements for
neutral pions are difficult, so these predictions may take time to be verified,
They are, nevertheless, very interesting.

Another remark is about the $e^+e^-$ annihilations, where two $W$ bosons are
produced. Such annihilations are being studied at LEP200 and as mentioned in
the previous section there is some controversy on the possible difference in
the Bose-Einstein correlations between identical pions originating from the
same $W$ boson as compared to pions originating from different $W$ bosons. In
the string model there is the unambiguous analysis for pairs of identical pions
from one string, i.e. from the decay of a single $W$ boson. For pions from two
strings, i.e. from two $W$ bosons, this analysis does not apply and another one
has not been proposed. In a recent report \cite{AN2} Andersson suggests that
"no cross talk is possible" for pions from different $W$ bosons, which would
mean no Bose-Einstein correlations. Of course, nobody doubts that pions are
bosons and that their state should have the corresponding symmetry with resect
to permutations of identical pions. However, in order to observe the effects
associated with the Bose-Einstein correlations since the GGLP paper, besides
the Bose-Einstein symmetry certain phase relationships are necessary, and these
may be absent for pairs of pions originating from different $W$ bosons. If this
string model prediction is taken seriously, it has interesting implications for
central heavy ions collisions. In such collisions so many strings are produced
that in a purely string picture two identical particles chosen at random are
very likely to originate from different strings and consequently would exhibit
no Bose-Einstein correlations. Since experimentally the Bose-Einstein
correlations in heavy ion collisions are quite strong, this means that the
state from which hadronization occurs is very different from a bunch of
independent strings.

\end{document}